%% file: main.tex
\documentclass[man, floatsintext]{apa7}
\input{0a-packages}

\title{Reading Speed, Image Quality Ratings, and Comfort Ratings\\in Augmented Reality}
\shorttitle{Reading in AR}
\authorsnames{Minjung Kim, Saeideh Ghahghaei Nezamabadi, Trisha Lian, Anand Singh}
\authorsaffiliations{{Meta Platforms Technologies, LLC.}}

\leftheader{Kim}
\abstract{\input{0b-abstract.tex}}

\authornote{
  \addORCIDlink{Minjung Kim}{0000-0002-3388-5947} \\E-mail: \url{mkim85@gmail.com}.  Web: \url{www.minjung.ca} The work was carried out while the author was an employee of Meta. Correspondence concerning this article should be addressed to Minjung Kim. }
     
\begin{document}
\maketitle

\input{1-introduction}
\input{2-background}
\input{3-methods}
\input{4-conditions}
\input{5-results}
\input{6-conclusion}
\input{acknowledgments}

\printbibliography

\end{document}

%% file: 0a-packages.tex
\usepackage[american]{babel}
\usepackage{url}
\usepackage{csquotes}
\usepackage[style=apa,sortcites=true,sorting=nyt,backend=biber]{biblatex}
\DeclareLabeldate{
  \field{date}
  \field{year}
  \field{eventdate}
  \field{origdate}
}
\DeclareLanguageMapping{american}{american-apa}
\usepackage{nameref}

\usepackage{float}
\usepackage{afterpage}
\usepackage{graphicx}
\graphicspath{ {./figs/} }
\usepackage{caption}

\makeatletter
\newcommand*{\flushqueue}{\bgroup
  \let\mylist=\@deferlist
  \gdef\@deferlist{}\par
  \loop\@next\mybox\mylist{}{\let\mybox=\voidb@x}%
  \ifvoid\mybox
  \else
    \ifnum\count\mybox<64
      \begin{figure}[ht!]
        \unvbox\mybox
      \end{figure}\par
    \else
      \begin{table}[ht!]
        \unvbox\mybox
      \end{table}\par
    \fi
  \repeat
\egroup}
\makeatother

\usepackage{setspace}
\usepackage{makecell}

\usepackage{multirow}

\usepackage[group-separator={,}]{siunitx}
\DeclareSIUnit{\nit}{
    \unit[per-mode=symbol]{\candela\per\meter\squared}
    }

\usepackage{amsmath}

\usepackage[T1]{fontenc}
\newcommand*{\fname}[1]{\texttt{#1}}

\usepackage[at]{easylist}
\AtBeginEnvironment{easylist}{\ListProperties(Start1=1)}

\usepackage{xspace} 

\newcommand{\projectname}{\textsl{Read-AR}\xspace}

%% file: 0b-abstract.tex
The rendering and display of text is a key use-case for augmented reality (AR). Here, we present the \projectname, a dataset of reading in AR, for which we collected over \num{11000}~reading speeds and almost \num{6000}~visual quality and comfort ratings across over \num{80}~different experiment conditions on the same experiment set-up. The consistent, controlled set-up enables the dataset to function as a reference for benchmarking the quality of different AR headset architectures. 

link to dataset on github: https://github.com/facebookresearch/ar-reading-dataset

%% file: 1-introduction.tex
\begin{figure}
    \centering
    \caption{Illustration of an augmented reality (AR) experience. The user of an AR headset would see an additive mixture of the AR content and the real world. Image composited in Photoshop \parencite{photoshop}, using stock images (\url{stock.adobe.com}).}
        
    \includegraphics[width=0.3\linewidth]{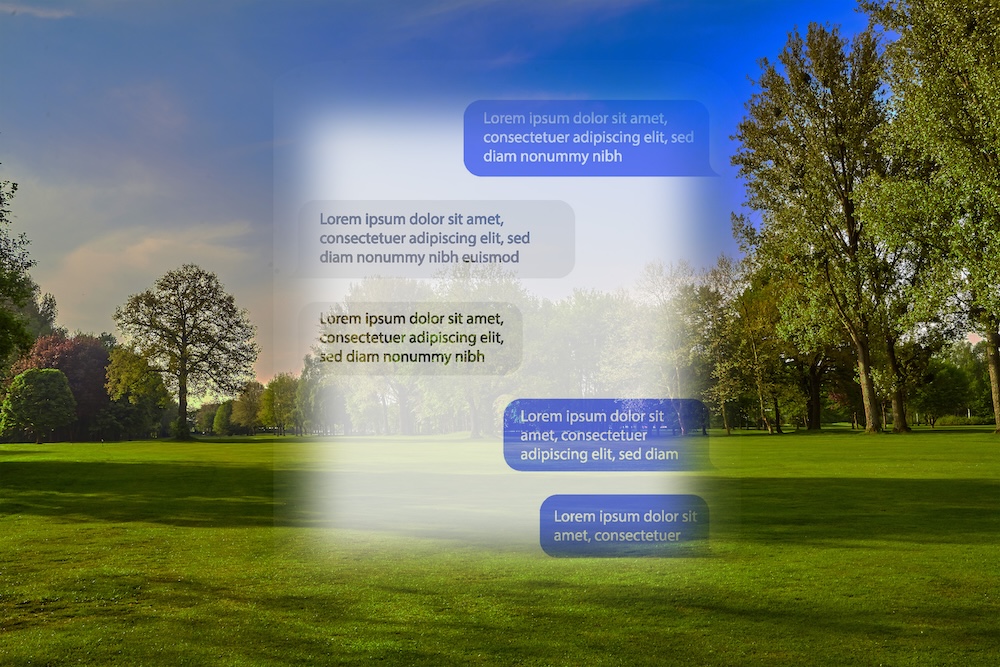}
    \includegraphics[width=0.3\linewidth]{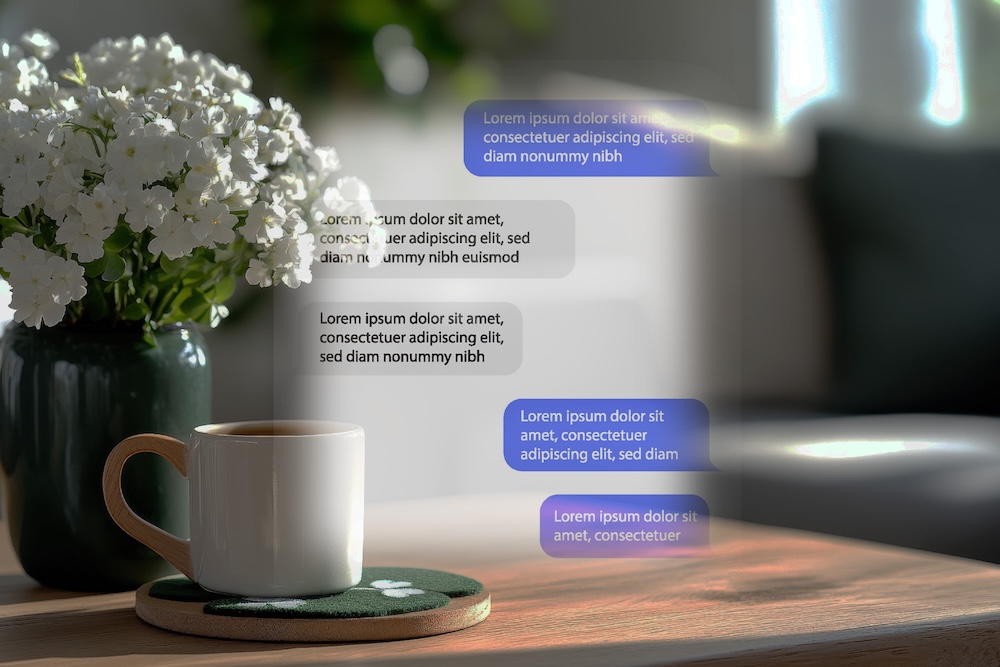}
    \vspace{1ex}

    \label{fig:ar-example}
\end{figure}

\label{sec:intro}
Near-eye displays and real-time head tracking have enabled immersive virtual and augmented experiences, called virtual reality (VR) and augmented reality (AR). Displaying high-resolution content can be difficult for VR and AR displays, due to low display resolution and loss of contrast from the optical design of the headset. AR displays also suffer from having to show digital content against the real world, whose luminance may be as high as \qty{10000}{\nit} in the sun, and which may also contain high-contrast textures that hinder visibility (Figure~\ref{fig:ar-example}).

Displaying text poses a particular technological challenge, as text is usually composed of small, thin lines, whose high-frequency signals are easily lost under low resolution and poor optical conditions. Displaying clear text is also a key use-case for AR, as text provides the user with an essential, low-friction method of interacting with the device. Thus, text-related metrics---such as reading speed, visual quality, and reading comfort---could be key behavioral and perceptual metrics for benchmarking the quality of different AR headset architectures. 

In this paper, we present \projectname, a large-scale dataset of reading measurements in an AR-like setting. The dataset consists of \num{11600} reading speeds, \num{5800} image quality ratings, \num{5800} comfort ratings, and \num{2175} freeform responses, measured across 108 participants with repetitions. The dataset was measured across multiple font sizes, display resolutions, optical quality, background luminance, and background textures (Table~\ref{tab:conds}). We hope that \projectname provides a reference point for evaluating different AR headset designs and that the dataset can be used to develop a model of reading in AR.

\flushqueue

%% file: 2-background.tex
\section{Background}\label{sec:bg}

Psychophysical studies of reading date back over 100 years \parencite{huey1898preliminary, quantz1897problems} and span wide range of topics, such as the typography of reading \parencite{tinker1963, legge2007, legge2011does, nanavati2005optimal}, eye movements and reading \parencite{rayner1998eye, bowers2017microsaccades}, the physiological basis of reading \parencite{yeatman2021reading}, reading for educational texts \parencite{grabinger1996designing, hughes2000typography}, reading in clinical populations \parencite{legge2007, o2000effect}, reading and discomfort \parencite{wilkins2021visual}, and letter identification in the periphery \parencite{pelli2007crowding}. 

For text legibility, the primary perceptual driver appears to be the font size, measured in degrees of visual angle. A unified definition of font size requires a weighting of different typographical parameters \parencite{wallace2022towards}, but is well-approximated by the x-height \parencite{legge2011does}. Metrics of legibility include maximum reading speed, critical print size, and reading acuity \parencite{legge2011does, mansfield1996psychophysics}. 

Early work on digital reading took place in the 1980s and 1990s, with the goal of understanding the requirements for text legibility on screens. Reading on paper was considered to be the gold standard, and replicating paper-like reading experience on a display was the goal \parencite{dillon1992reading}. Common concerns included eye strain, also called asthenopia or computer vision syndrome, with display resolution and frame rates often being the focus of the research. The field addressed questions such as the role of display resolution on reading speed \parencite{gould1987readinga, gould1987readingb, larimer2004hyperacuity} and the importance of antialiasing on text \parencite{gille1994gray}. Outcomes of the research included subpixel rendering and font hinting technology in the form of Microsoft ClearType \parencite{microsoft2001, microsoft2022} and to the design of the typefaces Georgia and Verdana, which were intended to be read on low-resolution screens \parencite{boyarski1998study, willharris2018}.

On standard modern displays, such as on a laptop, the visual quality of text is not generally limited by display resolution. For example, a 2021 MacBook Pro with Retina display has a native resolution of \qtyproduct{3024 x 1964}{px}. At a viewing distance of \qty{0.57}{m}, the angular resolution is \qty{101.8}{ppd} with a Nyquist rate of \qty{50.9}{cpd}, and above the visual acuity required for 20/20 vision. 

Rather, recent work on digital reading has investigated factors beyond the minimum requirement for legibility, such as how reading may take place on ``on the go'' on a smart watch or on an infotainment system of a car. For small, wearable displays, the size of the display restricts the number of letters per line width, which limits reading performance \parencite{atilgan2020reconciling, bernard2001examining}. Lower level factors such as type face, and higher level factors such as attention may also contribute, as users alternate between multiple tasks and engage in glancing or skimming \parencite{reimer2014assessing, beier2022readability, dobres2016utilising}. Other research has investigated how the aesthetics of typefaces may interact with the comfort of reading \parencite{pombo2025beautycomfort}; or how font type impacts crowding and reading  \parencite{beier2025applying}. 

In virtual reality, display's temporal dynamic such as display persistence could impact eye movements  \parencite{wu202268}. Display persistence is the illumination time within each frame. In virtual reality head mounted displays, low display persistence causes saccadic targeting errors. However, Wu and Murdison (2022)  did not find an effect on reading speed. It is not clear how the spatio-temporal dynamics of spatial attention in reading  \parencite{ghahghaei2013effects} may be impacted by the display's temporal dynamics. 

Other work has explored visual quality metrics tailored to text and reading, analogous to image quality metrics for natural images. These approaches capture a holistic judgment of quality even in cases where the text is legible and reading speed is optimal. For example, Zhang et al. \parencite{zhang2023textquality} trained a deep convolutional neural network on text classification and tuned it to text quality judgments. By processing an image of text, this metric incorporates factors such as resolution, blur, contrast, and text size, and predicts quality when user data is not available. Datasets like \projectname can be used as training data to further improve metric performance. 


\subsection{Reading in AR}
As AR headsets remain a fairly niche technology, there is little published work on the topic of reading in AR. Most of the research has focused on words or short sentences. Text is often used in experiments studying AR and additive displays, because it is both a demanding stimulus (because information is encoded at high spatial frequencies), yet also represents a plausible, and in fact common, use case for AR (as compared to other highly-sensitive stimuli, such as gratings, grids, and other geometric stimuli that are often used to study human vision \parencite{watson1983does}. 

Gabbard et al\parencite{gabbard2006effects} quantified the readability of individual characters, by asking participants to identify a single numeral embedded within a string of letters, as a function of properties such as background texture, luminance, and viewing distance. Similarly, Falk et al. \parencite{falk2021legibility}  quantified the readability of individual letters, by measuring visual search for a target letter, as a function of display and text properties such as polarity, luminance, and contrast. The second experiment measured reading speed and comprehension for 1, 2, 3 lines of text, of varying sizes and polarity. At one level higher, individual words have been used to study various aspects of visual perception. MacKenzie and Blanc-Goldhammer \parencite{blanc2018effects} measured reading speed as a function of additive contrast and other display properties, by asking participants to search for a target word within a set of 5 words presented using rapid serial visual presentation (RSVP). Falk et al. \parencite{falk2021legibility}  reported a second study, which quantified reading speed and comprehension for 1, 2, 3 lines of text, of varying sizes and polarity. 

Binocular differences need to be taken into account. Gurman, Spiegel and Rio  \parencite{Gurman2025VerticalBinocular} measured the time required for participants to fuse an AR image with vertical binocular misalignment, by asking them to report which of two individual words was a real word in English (with the other being a nonsense word). Spiegel and Erkelens \parencite{falk2021legibility} looked at binocular conflicts that are likely to impact reading performance. These conflicts may arise from mismatched vergence and accommodation cues—commonly referred to as vergence-accommodation conflicts (VAC) \parencite{spiegel2024vergence}—or from incorrect or missing occlusions \parencite{spiegel2025stereopsis}. Recent research has demonstrated that such conflicts can significantly impair the accurate discrimination of single letter orientation.

Lastly, a small number of studies have investigated reading over long durations – a few sentences to paragraphs. Erickson et al \parencite{erickson2021extended}  asked participants to read 4 paragraphs over a period of 1 minute, presented a Microsoft HoloLens AR HMD, and measured visual acuity, subjective usability, and preference as a function of text polarity, luminance, and background texture. Bilal Şimşek et al. \parencite{csimcsek2025examining}  measured reading comprehension during an AR-augmented multimedia experience, but used tablet-based passthrough rather than optical see-through AR, limiting generalizability to additive displays. 

For a more thorough review of studies on reading in AR, see Cauz, Clarinval and Dumas \parencite{cauz2024text} . 

To our knowledge, \projectname is the first dataset to offer comprehensive, standardized measurements of reading speed, image quality ratings, and comfort ratings across many different experiment conditions in AR.

\flushqueue

%% file: 3-methods.tex
\section{General Methods}\label{sec:methods}
All data were collected under Advara IRB oversight, Pro00073970 (PSIQ-2023: Perceptual Evaluation and Quality of Display System).

\subsection{Observers}
108 observers participated in data collection with informed consent. Five were Meta employees. The dataset excludes raw data from four observers, whose sessions were run with incorrect line spacing. 

\begin{figure}
    \centering
    \caption{Demographics of the observer pool.}
    \includegraphics{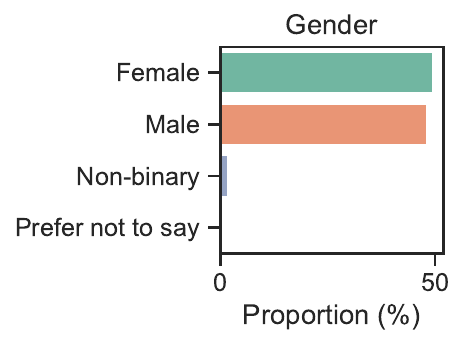} 
    \includegraphics{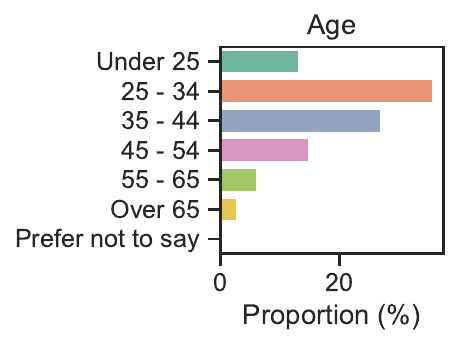}
    \\
    \includegraphics{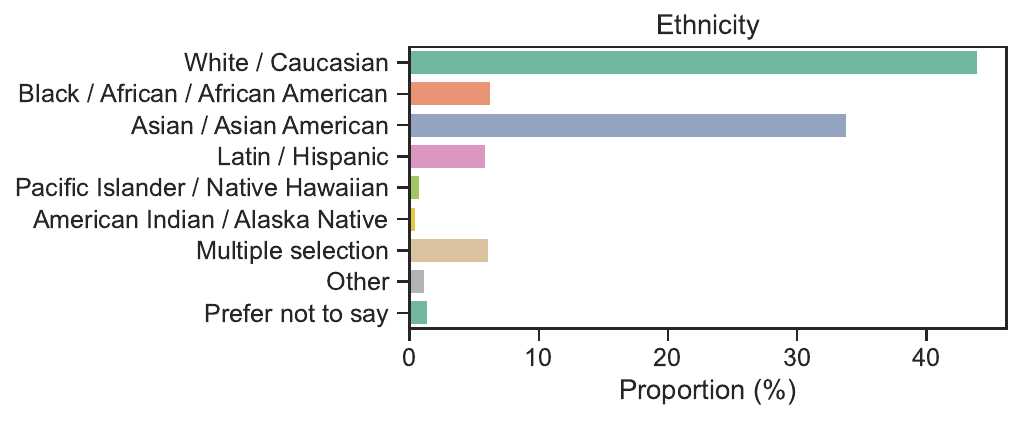}
    \label{fig:demographics}
\end{figure}

We provide information on participants based on the recruitment pool from 2019 to 2025 (Fig.~\ref{fig:demographics}). 

All observers had normal or corrected-to-normal visual acuity and normal stereoacuity, confirmed with the high-contrast Bailey-Lovie letter acuity chart \parencite{bailey1976new} and the Randot stereoacuity test \parencite{randot}. All observers described themselves as fluent in professional English, but were not subjected to a linguistic test to confirm.

\subsection{Apparatus}

\begin{figure}
\centering
\caption{Schematic of the AR stereoscope.}
\includegraphics{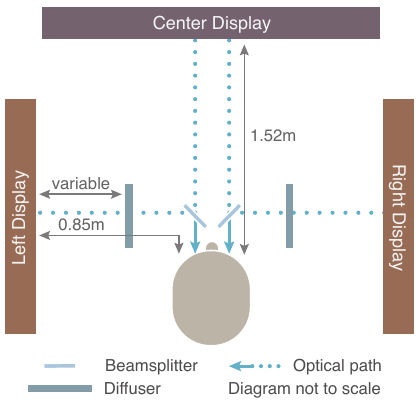}
\label{fig:ar-stereoscope}
\end{figure}

All stimuli were presented on a Wheatstone stereoscope \parencite{wheatstone1838stereoscope, wade2002charles}, modified to simulate an AR viewing environment. This AR stereoscope consisted of three monitors (left, right, and center), two diffusers (left, right), and two beamsplitters (left, right). See Figure~\ref{fig:ar-stereoscope}.

The left and right monitors \parencite[Eizo CG3145; native resolution: \qtyproduct{4096 x 2160}{px};][]{cg3145} were mounted at an optical distance of \qty{0.85}{\m} (\qty{1.18}{D}) from the observer's eyes, subtending \qtyproduct{47.1 x 24.8}{\degree} with an angular resolution of \qty{87.0}{ppd}. The center monitor was a DynaScan DS752LT5 \parencite[native resolution: \qtyproduct{3840 x 2160}{px};][]{ds752lt5}, placed at a distance of \qty{1.52}{\m} (\qty{0.66}{D}), subtending \qtyproduct{43.8 x 25.5}{\degree} with an angular resolution of \qty{87.7}{ppd}. 


The diffusers \parencite[\ang{0.5} Diffusing Angle Holographic Diffuser; ][]{diffuser05} were placed in front of the left and right monitors; the placement depended on the experiment condition. The diffusers scattered light from the monitors, blurring the displayed images. The amount of blur was quantified using ISO~12233, a method for characterizing the modulation transfer function (MTF) of an optical system \parencite{iso12233}. 

The beamsplitters \parencite[75R/25T Plate Beamsplitter;][]{beamsplitter75r25t} took the place of mirrors on a standard stereoscope, and were key to the AR simulation. The beamsplitters combined light rays such that the visual stimulus was an additive optical mixture of the content presented on the left and right (\qty{75}{\percent}) and in the center (\qty{25}{\percent}), as in an AR headset. 

All luminance and color calibration took place through the beamsplitters, meaning the spectrophotometer measurements accounted for the loss of light and for any shift in chromaticity. The whitepoint of all three monitors were set to a D65 metamer. The local dimming feature on the center display was disabled prior to data collection; the left and right displays had no such feature.

\subsection{Corpus}

The text content was extracted from \textsl{Anne of Green Gables} \parencite{montgomery1908anne}, a classic children's literature, chosen in part due to its reading level \parencite[Grade 5-8 reading level, or 970L; ][]{lexile_anne, lexile_chart} and its length (approximately \num{100000} words). The length of \textsl{Anne of Green Gables} provided sufficient content for observers to read new passages across multiple conditions and multiple sessions. By comparison, standard corpora for evaluating reading, such as the MNREAD \parencite{mnread} and the IReST \parencite{trauzettel2012standardized}, are short and would have required observers to read the same text repeatedly, possibly affecting reading times. 


\subsection{Stimuli}
The text rendering and presentation were controlled with EasyEyes \parencite{easyeyes}, an open source platform for conducting experiments on reading. It allowed us to set font size in degrees and take high-precision measurements of reading times.

We displayed test on the left and the right monitors, using  Optimistic Text \parencite{optimistic} with regular font weight. The text was white (\qty{500}{\nit}) and the background was nearly black (\qty{1.25}{\nit}). Due to the additive optics of the AR stereoscope, white text appeared to be an additive, transparent overlay against the center monitor, and the background was nearly invisible. 

The font size depended on the experiment condition. As each page had approximately 100 words, conditions with different font sizes subtended different angular size. For all font sizes, the text region was completely encompassed by the central display, to ensure that the luminance and texture of the background content contributed to the AR reading experience.

\subsection{Procedure}
Each experiment session consisted of three to four reading conditions, and each condition consisted of four reading trials. 

In each trial, observers silently read text from \textsl{Anne of Green Gables} \parencite{montgomery1908anne}, split into four pages of approximately 100 words each ($M=99.5, SD = 7.93$), which we controlled by setting EasyEyes to target 36.5 characters per line and 15.5 lines per page.

At the end of the trial, the observers were asked to presented with a multiple choice word recognition test, in which they identified the word that they saw in the four pages. The purpose of the word recognition test was to ensure that the observers were not skimming the text, and was a part of the ordinary reading task provided in EasyEyes. 

The observers also rated the text in terms of visual quality and comfort, using a Likert-like scale from 1 (not at all) to 6 (very much). We asked each question twice, with slightly different phrasing:

\begin{easylist}[enumerate]
\ListProperties(Margin1=1.75cm)
@ How much beauty did you feel from the letters in the text? (How much did you like the appearance of the text?)
@ How comfortable did you find the reading experience?
@ How much did you like the visual quality of the text?
@ How much did you like the comfort of the reading experience?
\vspace{2ex}
\end{easylist}

At the end of the condition, the observers provided responses to the following open-ended questions before proceeding:

\begin{easylist}[enumerate]
\ListProperties(Margin1=1.75cm)
@ Were there any words that you did not know?
@ Imagine this was the reading experience in a new product. How likely were you to recommend this product to a friend and why?
@ How would you describe your experience overall reading the passages?
\vspace{2ex}
\end{easylist}

Each session lasted 1.5 to 2 hours. Observers were allowed to participate in multiple sessions. 50 observers chose to return for additional sessions; observers participated in $M = 1.82$ $(SD = 1.30)$ sessions.

\flushqueue

%% file: 4-conditions.tex
\section{Experiment Conditions}
We manipulated different experiment parameters by changing settings on the AR stereoscope. We manipulated font size and display resolution by changing the left and right monitors, the optical quality by changing the placement of the diffusers, and the background luminance and texture by changing the image displayed on the center monitor. Overall, we tested 84 different combinations of font sizes, display resolutions, optical quality, background luminance, and background textures, with varied number of observers per condition ($M = 9.36$, $SD = 3.37$); see Table~\ref{tab:conds}. In the remainder of this section, we describe how we operationalized the definitions of the stimulus parameters and manipulated them.

\begin{table}
\renewcommand{\arraystretch}{2.25}
\centering
\caption{Experiment conditions.}
    \begin{tabular}{p{0.275\linewidth} p{0.225\linewidth} p{0.425\linewidth}}
    \Xhline{1.2pt}
    
    \thead{Parameter}                & \thead{Operationalized \\Definition} & \thead{Values}  \\
    
    \Xhline{0.5pt}
    
    \makecell{Font size}             & X-height                   & \qtylist{0.1; 0.172; 0.2; 0.3}{\degree}    \\
    \makecell{Display resolution}    & \makecell{Rendering resolution} & \makecell{\qty{21.75}{ppd}, \qty{29.0}{ppd}, \qty{43.5}{ppd}, \qty{87}{ppd}}\\
    \makecell{Optical quality}       & Diffuser placement         & \makecell{\qty{0.0}{\cm} (no diffuser), \qty{6.9}{\cm}, \\ \qty{9.4}{\cm}, \qty{14.4}{\cm}, \qty{16.9}{\cm}}\\
    \makecell{Background luminance}  & Additive contrast          & $5\!:\!1$, $3\!:\!1$, $\infty\!:\!1$       \\
    \makecell{Background texture} & Image clutter                 & Plain (no texture), low, medium         \\[2ex]
    
    \Xhline{1.2pt}
    
    \end{tabular}
\label{tab:conds}
\end{table}

\subsection{Font Size}
We standardized font sizes to x-height in degrees, treating an x-height of \ang{0.2} as the reference size at which an average adult can read printed text at maximum reading speed \parencite{legge2011does}. Here, printed text represents high-quality text rendering, without image quality loss due to low display resolution or poor optics. 

For Optimistic Text, the font size was 1.85 times the x-height, and we controlled the font size using EasyEyes on the left and right monitors. See Fig.~\ref{fig:xht-vs-ppd} for sample renderings.

\subsection{Display Resolution}
\begin{figure}
\centering
\caption{Sample font sizes and resolutions tested in the experiment.}
\includegraphics[]{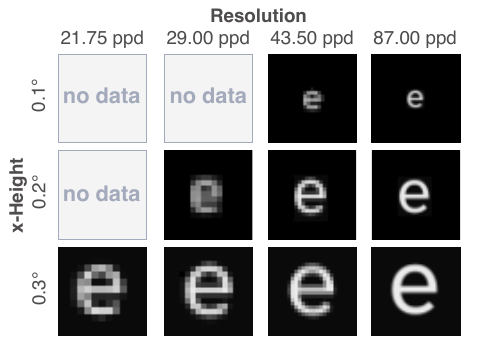}
\label{fig:xht-vs-ppd}
\end{figure}

We simulated AR headsets with display resolutions by setting the left and right monitors of the AR stereoscope to different resolutions. The native resolution of the AR stereoscope was \qtyproduct{4096 x 2160}{px}, or \unit{87}{ppd}, and was the upper limit of the resolution that we could test. We chose all other resolutions by dividing the native resolution with integers (\textsl{e.g.,} \qtyproduct{4096 x 2160}{px} $/\,2$ = \qtyproduct{2048 x 1080}{px}, or \qty{43.5}{ppd}). The integer divisor was critical for avoiding interpolation artifacts. See Fig.~\ref{fig:xht-vs-ppd} for sample renderings.

\subsection{Optical Quality}
\begin{figure}
    \centering
    \caption{MTF of the diffusers as a function of placement from the left and right monitors. HWHM is the half-width, half-maximum of the MTFs.}
    \includegraphics[width=0.9\linewidth]{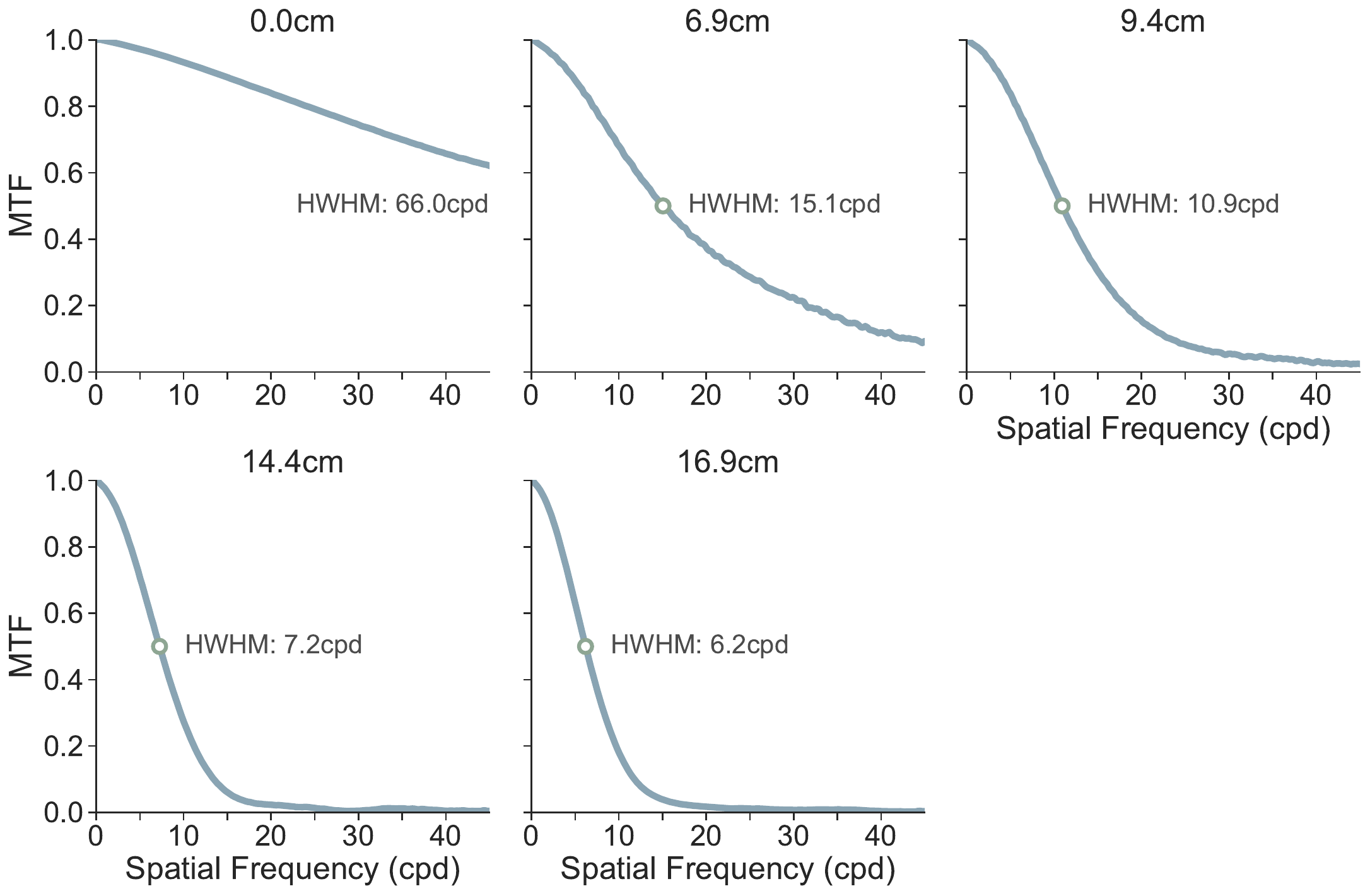}

    \label{fig:diffuser-mtf}
\end{figure}

We simulated AR headsets with different optical quality by placing the diffusers at different distances from the left and right monitors. The diffusers scattered light from the monitors, creating blur and lowering optical quality; larger distances resulted in worse optical quality \parencite[Figure~\ref{fig:diffuser-mtf}; ][]{iso12233}.

\subsection{Background Luminance}
The background luminance changes the adaptation state of the visual system, and therefore, its contrast sensitivity. This means that the legibility and image quality of text depend on both the text luminance and the background luminance.

In our study, we expressed the background luminance as \textsl{additive contrast}. Additive contrast is the AR analogue of contrast ratio, a commonly used metric in display engineering, and allowed us to incorporate the luminance contribution of the background while describing display performance. On a standard display, the contrast ratio $C$ is
\begin{align*}
\displaystyle
C &= \frac{L_{\max}}{L_{\min}}
\end{align*}

\noindent where $L_{\max}$ and $L_{\min}$ correspond to the maximum and the minimum luminance on the display. In our experiment, $L_{\max} = \qty{500}{\nit}$ and $L_{\min} = \qty{1.25}{\nit}$.

On an AR display, we calculate the addtivie contrast $C_\textrm{add}$, 
\begin{align*}
\displaystyle
C_\textrm{add} = \frac{L_{\max} + L_\textrm{bg}}{L_{\min} + L_\textrm{bg}}
\end{align*}

\noindent where $L_\textrm{bg}$ is the background luminance. For a homogeneous background, $L_\textrm{bg}$ is well-defined. For a complex scene, $L_\textrm{bg}$ represents the adaptation luminance, and could be the mean or the median luminance of the background image. In our study, we used the median. See Table~\ref{tab:add_cst} for full set of $L_{\max}$, $L_{\min}$, and $L_\textrm{bg}$ used.

\begin{table}
\renewcommand{\arraystretch}{1.5}
\centering
\caption{Additive contrast and the corresponding set of luminance used in the experiment.}.
    \begin{tabular}{cccc}
    \Xhline{1.2pt}
                & \multicolumn{3}{c}{\thead{Luminance}} \\ 
    \cline{2-4}
    \thead{Additive Contrast} & \thead{Max AR} & \thead{Min AR} & \thead{Background} \\
    
    \Xhline{0.75pt}
    
    $\infty:1$  & \multirow{3}{*}{\qty{500}{\nit}} & \multirow{3}{*}{\qty{1.25}{\nit}} & \qty{<0.1}{\nit}  \\  
    $5:1$       & & & \qty{123.4}{\nit} \\
    $3:1$       & & & \qty{248.1}{\nit} \\
    
    \Xhline{1.2pt}
    
    \end{tabular}

\label{tab:add_cst}
\end{table}

\subsection{Background Texture}
\begin{figure}
    \centering
    \caption{Stock images (\url{stock.adobe.com}) that are similar in appearance to the background images used in the experiment. Left: low clutter. Right: medium clutter.}
        
    \includegraphics[height=1.5in]{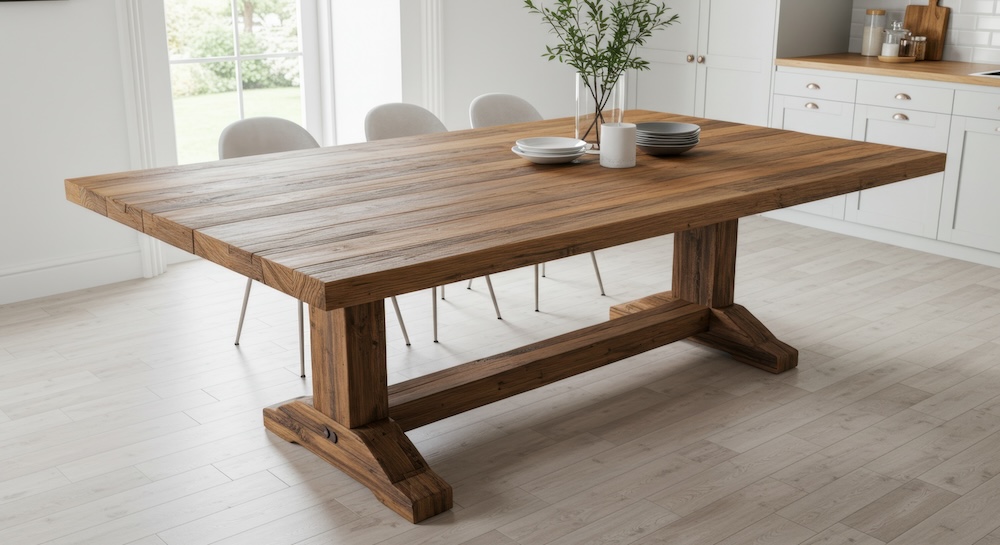}
    \includegraphics[height=1.5in]{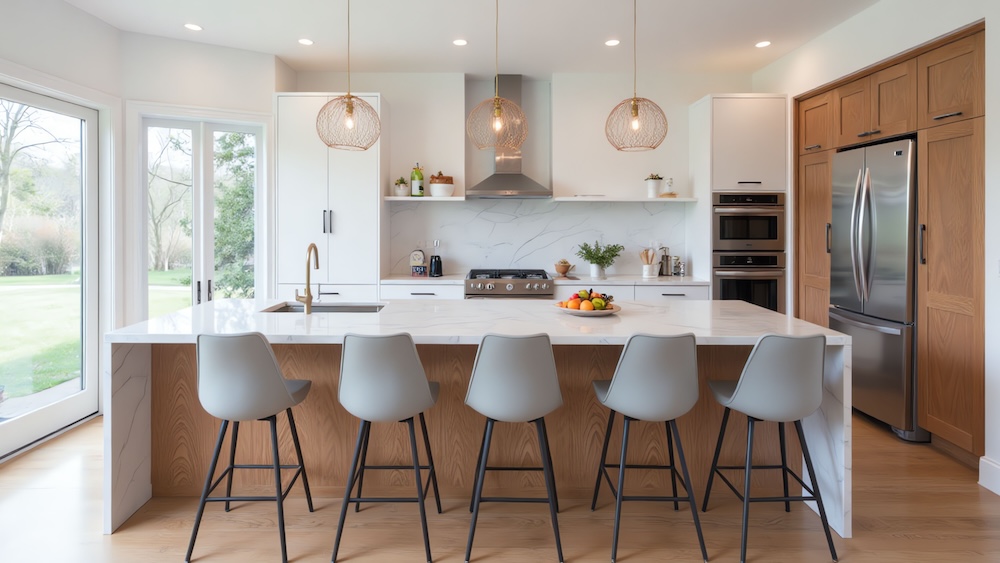}
    
    \label{fig:bg-texture}
\end{figure}

We considered three background textures: plain (no texture), low clutter and medium clutter. The plain background was only used with $\infty:1$ additive contrast. The low and medium clutter backgrounds were photographs that represented a typical home. We cannot provide the exact low and medium clutter images used in the experiment, but can provide a verbal description and stock images with similar appearance (Figure~\ref{fig:bg-texture}).

The low clutter image was of a living room. The centerpiece of the image was a large, wooden desk with a wooden chair. There were no objects on the desk. In the background, a cabinet with decorative wooden panels were visible, along with some house plants. On all wooden materials in the image, the wooden texture was visible, but low contrast and subtle. Text stimuli were presented against the wooden desk and the floor.

The medium clutter image was of a kitchen. In the center of the image, there was a kitchen island with two crockpots. and four wooden stools. Behind the kitchen island, cabinets and a window were visible. The cabinets were filled with cups and mugs. The window showed a peek into a brightly lit garden. Text stimuli were presented against one of the crockpots, the hexagonal tiling of the kitchen wall, and the window with a view of the garden.

\flushqueue

%% file: 5-results.tex
\section{The Dataset}
The \projectname dataset is available on the Meta Research GitHub repository:

\hspace{1cm}\url{https://github.com/facebookresearch/ar-reading-dataset}
    
\noindent The main dataset is \fname{data.csv}. The columns contain the independent and dependent variables. Each row corresponds to one condition by one observer. We provide additional supporting files, which provide additional helpful information. See Table~\ref{tab:dataset} for description of all files.

\begin{table}
\renewcommand{\arraystretch}{1.5}
\centering
\caption{Data files.}
    \begin{tabular}{l p{3.5in}}
    \Xhline{1.2pt}
    \thead{File Name} & \thead{Description} \\
    
    \Xhline{0.75pt}
    
    \fname{data.csv}  &  Main dataset. Includes condition IDs, experiment parameters, and all responses.\\  
    \fname{suppl-1-cond\_ids.csv} & Mapping between condition IDs and experiment parameters, as well as number of observers in that condition as an convenient overview.\\
    \fname{suppl-2-size\_spacing\_ppd.csv} & A table of font sizes and resolutions tested in the experiment. \\
    \fname{suppl-3a-mtf.csv} & Mapping between MTFs and diffuser placement for the conditions tested in the experiment. The MTFs are fitted based on MTF calibration data in \fname{suppl-3b-mtf\_calibration.csv}.\\
    \fname{suppl-3b-mtf\_calibration.csv} & Mapping between MTFs and diffuser placement based on calibration data  \parencite[ISO~12233; ][]{iso12233}.\\
    \fname{suppl-4-additive\_contrast} & Mapping between the additive contrast and the luminance values used in the experiment.\\
    \fname{suppl-5-rating\_freeform\_qns.csv} & Mapping between question headings in \fname{data.csv} and actual wording used.\\
    \Xhline{1.2pt}
    
    \end{tabular}

\label{tab:dataset}
\end{table}

\flushqueue

%% file: 6-conclusion.tex
\section{Conclusion}

We reviewed the unique challenges and of rendering legible and high quality text in AR devices. We also highlighted opportunities to apply our understanding of visual encoding processes and information processing that are unique to reading. We hope that the \projectname dataset can drive display development toward perceptually informed algorithms and technical requirements.

\flushqueue

%% file: acknowledgments.tex
\section{Acknowledgments}\label{sec:acknowledgments}
A project of this scope could not have taken place without the help and support of many people over several years. We are indebted to the following people for their contribution. \textbf{Apparatus:} Steven Smith, Dennis Pak, Xin Li, Will McCann. \textbf{Software, Easyeyes:} Denis Pelli, Nati Tsegaye, Gedion Alemayehu, August Burchell, Chongjun Liao, Yonathan Wagaye. \textbf{Software, Meta:} Sravya Motamarri, Natrayan Kuppusamy, Ali Yousefi. \textbf{Logistics:} Sara Kenley, Jacqueline Beyrouty, Jessica Grey, Eloise Moore, Kayla Garcia, John Hill, Katelyn Troastle. 
\textbf{Research assistance:} Cameron Wood, Michael Hicks, Avery Ao. \textbf{Helpful Discussion:} Denis Pelli, Maria Pombo, Anna Bruns. \textbf{Ethics:} Jeremy Johnsen. \textbf{Management:} Romain Bachy. 

\flushqueue